\documentclass[Conference]{IEEEtran}
\IEEEoverridecommandlockouts

 \usepackage{color}
\usepackage{graphicx}
\usepackage{amsmath}
\usepackage{amssymb}
\usepackage{caption}

\newcommand{\be}{\begin{equation}}
\newcommand{\ee}{\end{equation}}
\newcommand{\ba}{\begin{array}}
\newcommand{\ea}{\end{array}}

\newcommand{\non}{\nonumber}

\renewcommand{\thesubsubsection}{\thesubsection.\arabic{subsubsection}}

\title{Transmit Filter and Artificial Noise Design for Secure MIMO-OFDM Systems\thanks{This paper is supported by the Natural Science Foundation of
China (Grant No. 61671101 and 61601080).}}

\author{ \IEEEauthorblockN{Wenfei Liu$^{\dag}$, Ming Li$^{\dag}$,  Xiaowen Tian$^{\dag}$, Zihuan Wang$^{\dag}$, and Qian Liu$^{ \ddag}$
\vspace{-0.0 cm} }\\
\IEEEauthorblockA{$^{\dag}$School of Information and Communication Engineering   \\  Dalian University of Technology, Dalian, Liaoning 116024, China  \\
E-mail: \texttt{\{liuwenfei, tianxw, wangzihuan\}@mail.dlut.edu.cn, mli@dlut.edu.cn}}

\IEEEauthorblockA{$^{\ddag}$  School of Computer Science and Technology \\  Dalian University of Technology, Dalian, Liaoning 116024, China \\ E-mail: \texttt{qianliu@dlut.edu.cn}} }

\begin{document}

%\begin{spacing}{2.0}

\pagestyle{empty}

 \maketitle

\vspace{-0.0 cm}

\begin{abstract}

Physical layer security has been considered as an important security approach in wireless communications to protect  legitimate transmission from  passive eavesdroppers.
This paper investigates the physical layer security of a wireless multiple-input multiple-output (MIMO) orthogonal frequency division multiplexing (OFDM)
 communication system in the presence of a multiple-antenna eavesdropper.
We  first propose a transmit-filter-assisted secure MIMO-OFDM system which can destroy the orthogonality of eavesdropper's signals.
Our proposed transmit filter can disturb the reception of eavesdropper while maintaining the quality of legitimate transmission.
Then, we propose another artificial noise (AN)-assisted secure MIMO-OFDM system to further improve the security of the legitimate transmission.
The time-domain AN signal is designed to disturb the reception of eavesdropper while the  legitimate transmission will not be affected.
Simulation results are presented to demonstrate the security performance of the proposed transmit filter design and
AN-assisted scheme in the MIMO-OFDM system.

\end{abstract}
\vspace{-0.0 cm}

\section{Introduction}
With the rapid development of wireless communication systems,
security has become  a crucial issue in wireless networks due to  open broadcast nature.
Especially, when the important information is transmitting via the wireless medium, an
eavesdropper can also receive the signals, resulting in a great threat to
  the security of the wireless communication systems \cite{Yi 11}.
Therefore, it is essential for the legitimate transceiver to apply security transmission schemes.
Current security schemes rely on the conventional encryption mechanism.
However, the cryptography algorithms are
based on the computational hardness assumptions, which are
still mathematically unproven.
Physical layer security is now recognized as an alternative technique in
achieving security transmission, which exploits the physical characteristic of wireless channels.

The main principle of physical layer security is making use of the
randomness of the wireless channels to disturb eavesdropper's reception
without relying on the upper-layer \cite{A 14}.
Many initial physical layer security techniques are investigated in an
information-theoretic way. For example,
a seminal physical layer security work proposed by Wyner \cite{Wyner}.
Wyner proved that if the wiretap channel was worse than the legitimate channel,
then the communication of the legitimate receiver can achieve secure transmission.
Based on the pioneering studies of physical layer security,
many researches on secrecy capacity have been studied for different channel
models, such as Gaussian noise channel,
fading channel, Gaussian broadcast channel, etc. %\cite{Gaussian}-\cite{Korner}
Beamforming and power allocation techniques are also investigated to achieve
secrecy capacity  \cite{U}-\cite{WHM1}.
Artificial noise (AN) is another physical layer security technique.
AN is generated in the null space of the legitimate channel
and can be canceled out at the legitimate receiver but cannot be removed at the eavesdropper.
%To obtain AN, the transmitter is assumed to know the channel state information (CSI) of the
%legitimate channel in many wireless systems \cite{P AN}-\cite{WHM2}.

%%MIMO-OFDM
While most of the previous researches focus on information-theoretic analysis,
there is also a growing need to provide practical security solutions.
Recently, OFDM has been widely used in wireless communications
due to its high performance and low complexity, and
 the physical layer security in the application of OFDM systems
 is also well-noticed.
 Early information-theoretic based techniques utilized in OFDM systems are
 basically channel coding and cryptography \cite{OW 09}-\cite{en}.
In \cite{Q AN}, the authors proposed a time-domain AN
based physical layer security technology in OFDM systems.
In MIMO systems, transmit beamforming designs have been universally
 utilized by exploiting the spatial degrees of freedom to enhance the physical layer security \cite{Beam 2}-\cite{Beam 5}.
 In MIMO-OFDM systems, which have both the advantages of MIMO systems and OFDM systems,
 physical layer security are also  studied  \cite{MIMO}-\cite{MIMO1}.

Inspired by the above  studies, in this paper, we propose a
transmit-filter-assisted secure scheme and AN-assisted secure scheme in  MIMO-OFDM systems.
We assume that the transmitter has perfect knowledge of the  channel state information (CSI)
of legitimate channel.
Based on the diversity of the legitimate channel and wiretap channel,
the transmit filter is designed based on the legitimate CSI to deliberately destroy the orthogonality of the transmitted signals.
Only the legitimate receiver can recover the original signals
by making use of the minimum  mean square error (MMSE) filter to
suppress the inter-carrier interference (ICI).
Finally, we apply the AN approach to further improve the transmission security.

The following notation is used throughout this paper.
$(\cdot)^T$ represents transpose of a matrix.
$(\cdot)^H$ represents conjugate transpose of a matrix.
$\mathbb{E}\{\cdot \}$ denotes statistical expectation.
$Tr \{\mathbf{X} \}$ denotes the trace of matrix $\mathbf{X}$.
$\mathbb{C}^{M \times N}$ represents $M \times N$ matrices.
$\mathbf{I}_N$ represents an $N \times N$ identity matrix.
$Diag(x_1,\ldots,x_M)$ represents the $M \times M$ diagonal matrix
whose diagonal elements are $x_1$ to $x_M$.
$null(\cdot)$ represents the orthonormal bases for the null space of a matrix.
$blkdiag(\mathbf{A}_1,\mathbf{A}_2,\ldots,\mathbf{A}_N)$ denotes a block diagonal matrix
where the enclosed elements are diagonal blocks.
$\otimes$ is the Kronecker product. The function $min(\cdot,\cdot)$ returns the minimum
among the values enclosed in brackets.
$\|\cdot\|$ denotes the Enclidean norm of a vector.
$[\cdot]_{i,k}$ denotes the $(i,k)$th entry of a matrix.
$[\cdot]_k$ denotes the $k$th entry of a vector.

\section{System Model and Problem Formulation}
\label{sec: model}
%加filter系统图
\begin{center}
  \begin{figure}[!tp]  %% figure* 居中
  \begin{center}
    \vspace{-0.3cm}
        \includegraphics[width= 3.6 in]{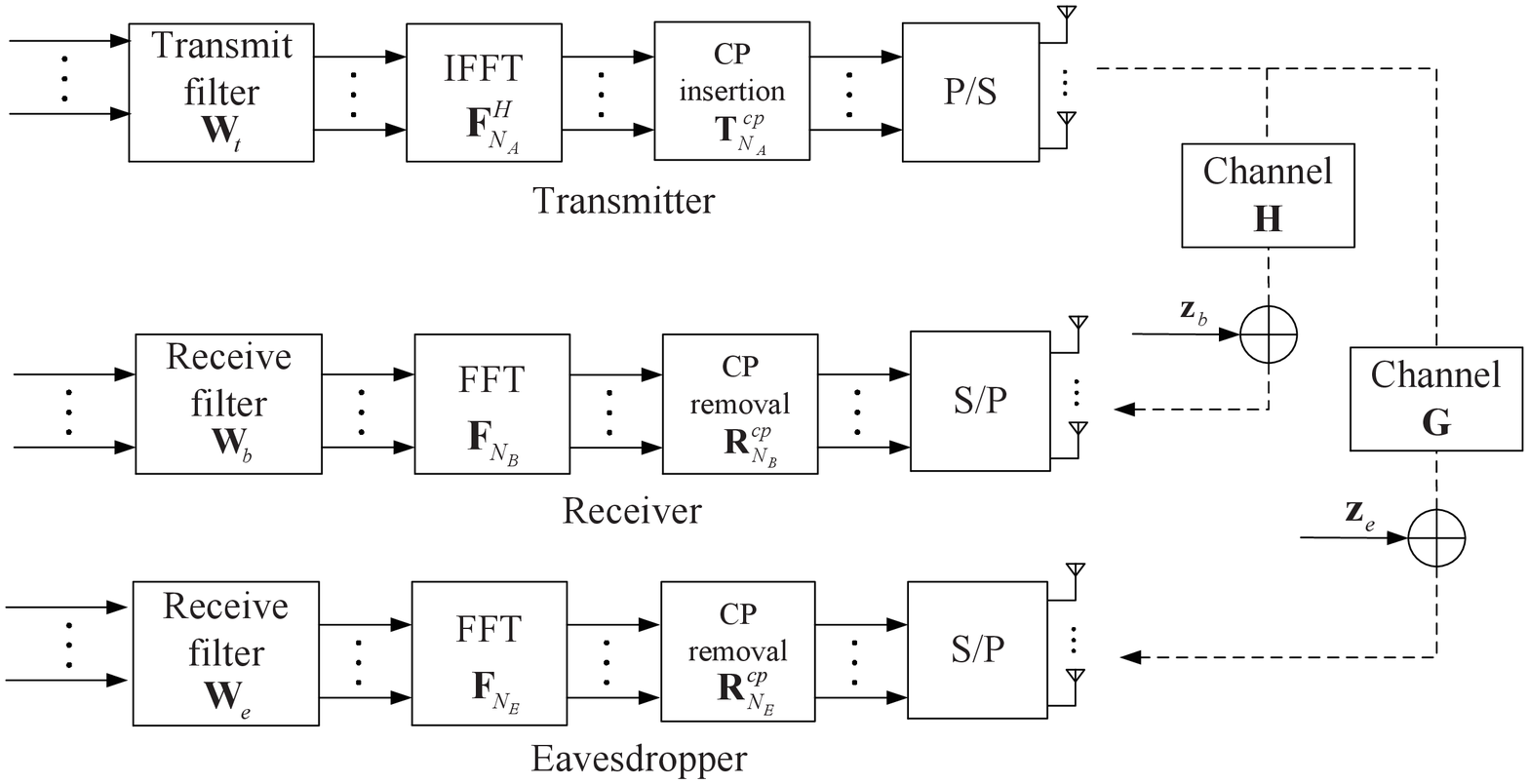}
           \vspace{-0.3cm}
    \caption{The filter-assisted secure MIMO-OFDM system.}
 \label{W1}
  \end{center}
\end{figure}
 \end{center}

We consider a MIMO-OFDM communication system  with one transmitter (Alice), one legitimate receiver (Bob),
and one eavesdropper (Eve).
The number of antennas at node $m$ $\in$ $\{A, B, E \}$ is $N_m$.

%\begin{center}
%\begin{figure}[!h]
%\begin{center}
%\includegraphics[width= 3.5 in]{alice.eps} \vspace{-0.3 cm}
%\caption{The wiretap MIMO-OFDM system with a multiple-antenna eavesdropper.}
%\label{trans}\vspace{-0.0 cm}
%\end{center}
%\end{figure}
%\end{center}
%

In order to clearly present our MIMO-OFDM system, we first
mathematically express the conventional single-input single-output (SISO)-OFDM transmission procedure.
Let $\mathbf{s}\triangleq[s_1,s_2,\ldots, s_N]^T$ denotes the transmitted signals,
where $N$ is the number of subcarriers.
$\mathbf{F}$ and $\mathbf{F}^{H}$ denote the $N\times N$ FFT and IFFT
matrices, respectively.
$\mathbf{T}_{cp}$  and $\mathbf{R}_{cp}$ %\triangleq[\mathbf{E}^{T}_{N_{cp}\times N}\;\mathbf{I}_N]^{T}
are the cyclic prefix (CP) insertion matrix and removal matrix, respectively.
%where $\mathbf{E}_{N_{cp}\times N}$ is the last $N_{cp}$ rows of the identity matrix $\mathbf{I}_N$.
The legitimate channel matrix $\mathbf{H}$ is a $ (N+N_{cp})\times (N+N_{cp})$ Toeplitz matrix
which can be expressed as
\vspace{0.2 cm}
\be  \mathbf{H}  =  \left[ \begin{array}{c c c c c}
 h(1) & 0 & 0 &\ldots & 0\\
\vdots &  h(1) & 0 & \ldots & 0 \\
h(L) & \ldots & \ddots & \ldots & \vdots\\
\vdots & \ddots & \ldots & \ddots & 0\\
0 & \ldots & h(L) &\ldots & h(1)
 \end{array}
\right] ,
\label{H} \vspace*{-0.0 cm} \ee
where $h(l), l\in{\{1,\ldots,L\}},$ is the channel impulse response from Alice to Bob,
 and $L<N_{cp}$ is the maximum delay spread.
At the receiver, frequency-domain signal of Bob can be expressed as
\vspace*{-0.0 cm}
%\begin{small}
\begin{eqnarray}
\mathbf{y}_b &=&\mathbf{F}\mathbf{R}_{cp} \mathbf{H} \mathbf{T}_{cp} \mathbf{F}^H \mathbf{P} \mathbf{s}+\mathbf{n}_b
\vspace*{-0.0 cm}
\end{eqnarray}
%\end{small}\\
where $\mathbf{n}_b\sim\mathcal{CN}(\mathbf{0},{\sigma}^2_n\mathbf{I}_N)$ is the additive white Gaussian noise (AWGN) vector at the legitimate receiver.

%In the system, Alice converts the frequency-domain symbols to
%time-domain symbols by the use of inverse fast Fourier transform (IFFT),
%and adds $N_{cp}$ length of cyclic prefix (CP) to the transmitted signals.
%Then, the transmitted signals are processed by a parallel-to-serial (P/S) converter
%to output the streams through a multipath Rayleigh fading channel.
%At the receiver, transmitted signals are processed by the serial-to-parallel (S/P) converter.
%After CP removal, the time-domain signals are transformed to the frequency-domain by FFT.
%

After explicitly explaining the mathematical model of the conventional SISO-OFDM system,
we turn to present the wiretap MIMO-OFDM system model as shown in Fig. \ref{W1}.
As for a MIMO-OFDM system, the original signal is firstly processed by a transmit beamformer/filter $\mathbf{W}_t$, and then
transformed to the time domain by IFFT.
After CP insertion, the transmitted  signal arrives at the receiver through
the multipath Rayleigh fading channel. Then, Bob and Eve will remove CP
and convert it to the frequency-domain signal by FFT. After that both Bob and Eve use receive
filters to recover the original signals.

To mathematically express the MIMO-OFDM system,
let $\mathbf{s}\triangleq[s_{1,1},\ldots,s_{1,N},\ldots,s_{N_s,1},\ldots,s_{N_s,N}]^T$
represent the symbol vector to be transmitted via the corresponding subcarriers,
$\mathbb{E}\{\|s_{i,j}\|^2\}=1$,
$N$ is the number of orthogonal subcarriers,
$N_s$ is the number of data streams of per subcarrier with $N_s \leq N_A$.
Let $\mathbf{F}_{N_m}=\mathbf{I}_{N_m}\otimes \mathbf{F} \in \mathbb{C}^{N_mN \times N_mN}$
denote the FFT operation at a node with $N_m$ antennas.
$\mathbf{F}_{N_B} = blkdiag(\mathbf{F},\mathbf{F},...,\mathbf{F})$ is the FFT matrix and
$\mathbf{F}_{N_A}^H = blkdiag(\mathbf{F}^H,\mathbf{F}^H,...,\mathbf{F}^H)$ is the IFFT matrix.
The matrix for CP insertion
is denoted by $\mathbf{T}_{N_A}^{cp}\in \mathbb{C}^{N_A(N+N_{cp})\times N_AN}$,
where $\mathbf{T}_{N_A}^{cp} = blkdiag (\mathbf{T}_{cp},...,\mathbf{T}_{cp})$.
$\mathbf{R}_{N_B}^{cp}\in \mathbb{C}^{N_B(N+N_{cp})\times N_BN}$ is the CP removal matrix, and
 $\mathbf{R}_{N_B}^{cp} = blkdiag (\mathbf{R}_{cp},...,\mathbf{R}_{cp})$.
 $\mathbf{W}_t \in \mathbb{C}^{N_AN \times N_sN}$ is the transmit beamformer/filter
 which has a power constraint, i.e. $Tr(\mathbf{W}_t \mathbf{W}_t^H) \leq P_t$,
 where $P_t$ is the total transmit power.
$\mathbf{H}\in\mathbb{C}^{N_B(N+N_{cp})\times N_A(N+N_{cp})}$ is the
Alice-to-Bob MIMO multipath channel matrix which is constructed by
\be
\mathbf{H}= \left[ \begin{array}{c c c}
\mathbf{H}_{1,1} & \ldots & \mathbf{H}_{1,N_A}\\
\vdots  & \ddots  & \vdots\\
\mathbf{H}_{N_B,1} &  \ldots & \mathbf{H}_{N_B,N_A}
 \end{array}\right] ,
 \label{HH}
\ee
$\mathbf{H}_{k,i}\in\mathbb{C}^ {(N+N_{cp}) \times (N+N_{cp})}$ is the channel impulse response between
the $i$th antenna of Alice and the $k$th antenna of Bob,
which has a form of
\vspace*{-0.0 cm}
\be  \mathbf{H}_{k,i}  =  \left[ \begin{array}{c c c c c}
 h_{k,i}(1) & 0 & 0 &\ldots & 0\\
\vdots &  h_{k,i}(1) & 0 & \ldots & 0 \\
h_{k,i}(L) & \ldots & \ddots & \ldots & \vdots\\
\vdots & \ddots & \ldots & \ddots & 0\\
0 & \ldots & h_{k,i}(L) &\ldots & h_{k,i}(1)
 \end{array}
\right].
\label{H}
\ee
The received frequency-domain signal at all antennas of the legitimate receiver can be expressed as
%\begin{small}
\begin{eqnarray}
\mathbf{y}_b &=&\mathbf{F}_{N_B} \mathbf{R}_{N_B}^{cp} \mathbf{H} \mathbf{T}_{N_A}^{cp}\mathbf{F}_{N_A}^H \mathbf{W}_t \mathbf{s}+\mathbf{z}_b \non \\
&=&  \widetilde{\mathbf{H}} \mathbf{W}_t \mathbf{s}+\mathbf{z}_b,
\end{eqnarray}
%\end{small}
where $\mathbf{y}_b \in \mathbb{C}^{N_BN \times 1}$ is the received signal vector at Bob,
$\mathbf{z}_b\sim\mathcal{CN}(\mathbf{0},{\sigma}^2_n\mathbf{I}_{N_BN})$ is the AWGN vector at Bob,
the effective channel matrix
%\begin{small}
\begin{eqnarray}
\widetilde{\mathbf{H}}& \triangleq & \mathbf{F}_{N_B} \mathbf{R}_{N_B}^{cp} \mathbf{H} \mathbf{T}_{N_A}^{cp}\mathbf{F}_{N_A}^H  \non \\
&=& blkdiag (\tilde{\mathbf{H}}_1, \tilde{\mathbf{H}}_2,\ldots, \tilde{\mathbf{H}}_N)
\end{eqnarray}
%\end{small}\\
is block-diagonal, where $\tilde{\mathbf{H}}_n\in\mathbb{C}^{N_B\times N_A}$
represents the frequency-domain channel matrix of the $n$th subcarrier
from the transmitter to the legitimate receiver.

The Alice-to-Eve channel matrix $\mathbf{G}$, whose size is
$N_E(N+N_{cp})\times N_A(N+N_{cp})$, has similar constructure as (\ref{HH}),
and the received signals at Eve can be given as
%\begin{small}
\begin{eqnarray}
\mathbf{y}_e &=&\mathbf{F}_{N_E} \mathbf{R}_{N_E}^{cp} \mathbf{G} \mathbf{T}_{N_A}^{cp}\mathbf{F}_{N_A}^H \mathbf{W}_t \mathbf{s}+\mathbf{z}_e \non \\
&=&  \widetilde{\mathbf{G}} \mathbf{W}_t  \mathbf{s}+\mathbf{z}_e,
\end{eqnarray}
%\end{small}
where $\mathbf{y}_e \in \mathbb{C}^{N_EN \times 1}$ is the received signal vector at Eve.
$\mathbf{F}_{N_E} \in \mathbb{C}^{N_EN \times N_EN }$ is the FFT matrix of Eve.
$\mathbf{R}_{N_E}^{cp} \in \mathbb{C} ^ {N_EN \times N_E(N+N_{cp})}$ is the CP removal matrix performed at Eve.
$\mathbf{z}_e\sim\mathcal{CN}(\mathbf{0},{\sigma}^2_n\mathbf{I}_{N_EN})$ is the AWGN vector at Eve.
The equivalent channel matrix
%\begin{small}
\begin{eqnarray}
 \widetilde{\mathbf{G}} &\triangleq & \mathbf{F}_{N_E} \mathbf{R}_{N_E}^{cp} \mathbf{G} \mathbf{T}_{N_A}^{cp}\mathbf{F}_{N_A}^H   \non \\
&=& blkdiag(\tilde{\mathbf{G}}_1,\tilde{\mathbf{G}}_2,\ldots,\tilde{\mathbf{G}}_N)
\end{eqnarray}
%\end{small}
 is also block-diagonal, where $\tilde{\mathbf{G}}_n\in\mathbb{C}^{N_E\times N_A}$
represents the frequency-domain channel matrix of the $n$th subcarrier
from the transmitter to the eavesdropper.

After describing the above system model,
we can see that  the equivalent channel matrix $\tilde{\mathbf{H}}$ and $\tilde{\mathbf{G}}$ have similar structures.
This will threaten the MIMO-OFDM systems because Eve can recover the transmitted signals after
some demodulation processes.
In order to prevent eavesdropping, we propose a transmit-filter-assisted secure MIMO-OFDM
technique to maintain the communication security.
The transmit filter aims to disturb the received signal of Eve while maintaining  the
Alice-to-Bob transmission,
as will be presented in Section \ref{sec: filter}.

\vspace{0.5 cm}
\section{Transmit Filter Design in MIMO-OFDM System}
\label{sec: filter}
This section is devoted to design the transmit filter in secure MIMO-OFDM system to
disturb the reception of Eve. % while not influence the reception of Bob.
We assume that Alice knows the CSI of the Alice-to-Bob channel link
while Eve does not.
Based on the diversity between the legitimate channel and wiretap channel,
Alice designs the transmit filter to deliberately
destroy the orthogonality of the transmitted signals.
Both Bob and Eve adopt the MMSE filter to suppress the  ICI.
However, Eve cannot recover the destroyed signals as what Bob does and
then the transmission security can be maintained.

After employing the transmit filter in the frequency-domain,
the received MIMO-OFDM signals at Bob and Eve are not orthogonal,
i.e. $\widetilde{\mathbf{H}}\mathbf{W}_t$ and $\widetilde{\mathbf{G}}\mathbf{W}_t$ are not diagonal matrixes.
At the receiver, Bob uses the MMSE receive filter to suppress the inter-subcarrier interference.
The recovered signal can be expressed as
\be \hat{\mathbf{s}}_b={\mathbf{W}^H _b}\mathbf{y}_b, \label{sb}  \ee
where ${\mathbf{W}_b}$ is the MMSE receive filter at Bob which
 is well known as
\begin{small}
\begin{eqnarray}
\begin{split}
\mathbf{W}_b =&(\widetilde{\mathbf{H}} \mathbf{W}_t \mathbf{W}_t^H \widetilde{\mathbf{H}}^H + {\sigma} ^2_{z_b} \mathbf{I} )^{-1} \widetilde{\mathbf{H}} \mathbf{W}_t  \\
=& (\mathbf{F}_{N_B} \mathbf{R}_{N_B}^{cp} \mathbf{H} \mathbf{T}_{N_A}^{cp}\mathbf{F}_{N_A}^H \mathbf{W}_t
\mathbf{W}_t^H \mathbf{F}_{N_A}{\mathbf{T}_{N_A}^{cp}}^H \mathbf{H}^H   \\
 & {\mathbf{R}_{N_B}^{cp}}^H
 \mathbf{F}_{N_B}^H + {\sigma} ^2_{z_b} \mathbf{I} )^{-1}  \mathbf{F}_{N_B} \mathbf{R}_{N_B}^{cp} \mathbf{H} \mathbf{T}_{N_A}^{cp}\mathbf{F}_{N_A}^H \mathbf{W}_t.
 \label{WB}
 \end{split}
\end{eqnarray}
\end{small}

Then, with the  above MMSE receive filter $\mathbf{W}_b$,
the mean squared error (MSE) at Bob is
\begin{small}
\begin{eqnarray}
\begin{split}
\Gamma_b =&  \mathbb{E } \{ \|  \hat{\mathbf{s}}_b  - \mathbf{s} \| ^2   \}   \\
=&  \mathbb{E }\{ \|  \mathbf{W}^H _b \mathbf{y}_b - \mathbf{s} \| ^2 \}    \\
%=&\mathbb{ E} { \{ {( \mathbf{W}^H _b \mathbf{y}_b - \mathbf{s})}^H  ( \mathbf{W}^H _b \mathbf{y}_b - \mathbf{s} )  \} }    \\
%=& \mathbb{E} \{ ((\widetilde{\mathbf{H}} \mathbf{W}_t \mathbf{s}+\mathbf{z}_b )^H  \mathbf{W}_b - \mathbf{s}^H )
% (  \mathbf{W}^H _b (\widetilde{\mathbf{H}} \mathbf{W}_t \mathbf{s}+\mathbf{z}_b )-\mathbf{s} ) \}  \non \\
=& Tr \{  \mathbf{W}_b {\mathbf{W}_b}^H  \widetilde{\mathbf{H}} \mathbf{W}_t  {\mathbf{W}_t }^H  {\widetilde{\mathbf{H}} }^H\}
 + {\sigma}^2_{z_b} Tr\{  \mathbf{W}_b {\mathbf{W}_b}^H \}   \\
 & -Tr\{  {\mathbf{W}_b}^H  \widetilde{\mathbf{H}} \mathbf{W}_t  \}
  - Tr \{  {\mathbf{W}_t} ^H  {\widetilde{\mathbf{H}}} ^H   \mathbf{W}_b \} +N_sN.
 \label{MSE1}
 \end{split}
\end{eqnarray}
\end{small}

Submiting the MMSE receive filter (\ref{WB}) in (\ref{MSE1}), we can rewrite the MSE at Bob as
\begin{small}
\begin{eqnarray}
\begin{split}
\Gamma_b =& Tr{\{ ( \widetilde{\mathbf{H}} \mathbf{W}_t  {\mathbf{W}_t}^H  {\mathbf{H}}^H +  {\sigma}^2_{z_b}  \mathbf{I})^{-1} \widetilde{\mathbf{H}} \mathbf{W}_t  {\mathbf{W}_t}^H  {\widetilde{\mathbf{H}}}^H \}}^2   \\
&  + {\sigma}^2_{z_b} Tr\{ ( \widetilde{\mathbf{H}} \mathbf{W}_t  {\mathbf{W}_t}^H  {\widetilde{\mathbf{H}}}^H +  {\sigma}^2_{z_b}  \mathbf{I})^{-1} \widetilde{\mathbf{H}} \mathbf{W}_t  {\mathbf{W}_t}^H  {\mathbf{H}}^H    \\
& ( \widetilde{\mathbf{H}} \mathbf{W}_t  {\mathbf{W}_t}^H  {\mathbf{H}}^H +  {\sigma}^2_{z_b}   \mathbf{I})^{-1} \}
   -2 Tr\{(\widetilde{\mathbf{H}} \mathbf{W}_t {\mathbf{W}_t}^H  {\mathbf{H}}^H    \\
& +{\sigma}^2_{z_b}   \mathbf{I})^{-1}   \widetilde{\mathbf{H}} \mathbf{W}_t  {\mathbf{W}_t}^H  {\widetilde{\mathbf{H}}}^H \} + N_sN.
 \label{MSE}
 \end{split}
\end{eqnarray}
\end{small}

Our purpose is to design the transmit filter $\mathbf{W}_t$  which minimizes the MSE ($\Gamma_b$).
Obviously, this problem is too difficult to be solved.
Therefore, we turn to find a suboptimal solution to design $\mathbf{W}_t$.
Since the effective channel $ \widetilde{\mathbf{H}}$ is available to Alice,
we can  make the  singular value decomposition (SVD) of $ \widetilde{\mathbf{H}}$  as
$\widetilde{\mathbf{H}}= \mathbf{U} \mathbf{\Sigma} \mathbf{V}^H$.
The columns of $\mathbf{U} \in \mathbb{C}^{N_BN \times N_BN}$ are the left-singular vector of $\widetilde{\mathbf{H}}$
 and the columns of $\mathbf{V} \in \mathbb{C}^{N_AN \times N_AN}$ are the right-singular vector of $\widetilde{\mathbf{H}}$.
$ \mathbf{\Sigma} = Diag(\sigma_1,\ldots,\sigma_\alpha) \in \mathbb{C}^{N_BN \times N_AN}$ contains the singular values,
 where $\alpha=min (  N_AN, N_BN)$, and $\sigma_1 \geq \cdots \geq \sigma_\alpha$.
To transmit $N_sN$ symbols, we simply select the first $N_sN$ columns of $\mathbf{V}$ as $\mathbf{V}_t$.

Then, the term $\widetilde{\mathbf{H}}\mathbf{W}_t  {\mathbf{W}_t}^H  {\widetilde{\mathbf{H}}}^H  $  in (\ref{MSE})
can be rewritten as
\be
\widetilde{\mathbf{H}}\mathbf{W}_t  {\mathbf{W}_t}^H  {\widetilde{\mathbf{H}}}^H = \mathbf{U} \mathbf{\Sigma} \mathbf{V}_t^H  \mathbf{W}_t  {\mathbf{W}_t}^H \mathbf{V}_t \mathbf{\Sigma}^H {\mathbf{U}}^H.
\label{SVD}
\ee
For simplification  design purpose, let
\be \mathbf{W}_t =  \mathbf{V}_t  \mathbf{P} \ee
be the transmit filter design,
where  $\mathbf{P}  $ denotes the transmit power matrix and  $\mathbf{P} = Diag(\sqrt{p_1},\ldots,\sqrt{p_{N_sN}}) $.
The equation (\ref{SVD}) can be further simplified as
\be \widetilde{\mathbf{H}} \mathbf{W}_t  {\mathbf{W}_t}^H  {\widetilde{\mathbf{H}}}^H = \mathbf{U} \mathbf{\Lambda} {\mathbf{U}}^H, \label{aaa} \ee
where $ \mathbf{\Lambda} = \mathbf{\Sigma}  \mathbf{P}^2 {\mathbf{\Sigma} }^H = Diag(\lambda_1,\ldots,\lambda_{N_sN}) $.
Substituting (\ref{aaa}) in the original objective function (\ref{MSE}),  $\Gamma_b$ can be further expressed as
\vspace*{-0.0 cm}
%\begin{small}
\begin{eqnarray}
\begin{split} %公式换行
\Gamma_b=& Tr \{ ( (\mathbf{\Lambda} + {\sigma}^2_{z_b}  \mathbf{I})^{-1}\mathbf{ \Lambda} )^2 \}
      + {\sigma}^2_n Tr \{  (\mathbf{\Lambda} + {\sigma}^2_{z_b} \mathbf{I})^{-1} \mathbf{\Lambda} \\
      & (\mathbf{\Lambda} + {\sigma}^2_{z_b} \mathbf{I})^{-1} \}
      -2 Tr\{  (\mathbf{\Lambda} + {\sigma}^2_{Z_b}  \mathbf{I})^{-1}\mathbf{ \Lambda}  \} +N_sN   \\
     =& \sum_{i=1}^{N_sN} \frac{\sigma^2_{z_b}}{\sigma^2_i p_i +\sigma^2_{z_b}  }.
\label{lambda}
 \end{split}
\end{eqnarray}
%\end{small}

After the above equation transform,
we turn to solve the power allocation problem for subcarriers
instead of the original optimization problem.
The final transmit filter design problem can be expressed as
%\begin{small}
\be
\begin{split}
min & \; \sum_{i=1}^{N_sN} \frac{\sigma^2_{z_b}}{\sigma^2_i p_i +\sigma^2_{z_b}  }\\
  s.t. & \; \sum_{i=1}^{N_sN}p_i \leq P_t,\\
  &\;  p_i \geq 0, i=1, \ldots, N_sN.
 \label{MIN}
   \end{split}
\ee
%\end{small}

This power allocation problem can be easily solved by some existing solvers
such as CVX solvers. After working out the power allocation of different subcarriers,
we can obtain the transmit filter  $\mathbf{W}_t =  \mathbf{V}_t  \mathbf{P}$.

%Eve
As for Eve, she  also uses MMSE receive filter to recover the transmit signals.
Similarly as Bob, the recovered signal of Eve can be expressed as
\be\hat{\mathbf{s}}_e={\mathbf{W}^H _e}\mathbf{y}_e , \ee
where $\mathbf{W}_e$ is the MMSE receive filter which has a  similar form as (\ref{WB})
\begin{small}
\begin{eqnarray}
\begin{split}
\mathbf{W}_e =&(\widetilde{\mathbf{G}} \mathbf{W}_t \mathbf{W}_t^H \widetilde{\mathbf{G}}^H + {\sigma} ^2_{z_e} \mathbf{I} )^{-1} \widetilde{\mathbf{G}} \mathbf{W}_t   \\
=& (\mathbf{F}_{N_E} \mathbf{R}_{N_E}^{cp} \mathbf{G} \mathbf{T}_{N_A}^{cp}\mathbf{F}_{N_A}^H \mathbf{W}_t
\mathbf{W}_t^H \mathbf{F}_{N_A}{\mathbf{T}_{N_A}^{cp}}^H \mathbf{H}^H   \\
&  {\mathbf{R}_{N_E}^{cp}}^H
 \mathbf{F}_{N_E}^H + {\sigma} ^2_{z_e} \mathbf{I} )^{-1} \mathbf{F}_{N_E} \mathbf{R}_{N_E}^{cp} \mathbf{G} \mathbf{T}_{N_A}^{cp}\mathbf{F}_{N_A}^H \mathbf{W}_t.
 \label{We}
 \end{split}
\end{eqnarray}
\end{small}
The MSE at Eve is
\begin{small}
\begin{eqnarray}
\begin{split}
\Gamma_e =& Tr \{  \mathbf{W}_e {\mathbf{W}_e}^H  \widetilde{\mathbf{G}} \mathbf{W}_t  {\mathbf{W}_t }^H  {\widetilde{\mathbf{G}} }^H\}
 + {\sigma}^2_{z_e} Tr\{  \mathbf{W}_e {\mathbf{W}_e}^H \} \\
 & -Tr\{  {\mathbf{W}_e}^H  \widetilde{\mathbf{G}} \mathbf{W}_t  \}
  - Tr \{  {\mathbf{W}_t} ^H  {\widetilde{\mathbf{G}}} ^H   \mathbf{W}_e \} +N_sN.
 \label{MSEeve}
 \end{split}
\end{eqnarray}
\end{small}

Based on the fact that the transmit filter $\mathbf{W}_t$ is designed to minimize $\Gamma_b$
by the use of  legitimate CSI $\mathbf{H}$, Bob can efficiently recover the original signals
by employing the receive filter. However, due to the diversity between the wiretap channel $\mathbf{G}$
and the legitimate channel  $\mathbf{H}$, the MSE at Eve $\Gamma_e$ is much higher than $\Gamma_b$.
Therefore, after applying the transmit filter,
the received signals at Eve will be disturbed and the security will be enhanced.

\section{Artificial Noise Design in MIMO-OFDM System }
\label{sec: AN}

After developing  the transmit-filter-assisted secure scheme,
we find that the transformed power allocation problem in section \ref{sec: filter} is intriguing.
And it is sensible to adopt another AN-assisted scheme utilizing the residual power to further guarantee the
security of the transmission.

Let $\gamma_b$ denote the MSE threshold of Alice-to-Bob transmission.
When $\Gamma_b \leq \gamma_b$, the quality of legitimate transmission can be maintained
to a certain extend according to $\gamma_b$.
Let $P_t$ represent the total transmit power.
Denote
\begin{small} $P_c= \sum_{i=1}^{N_sN} p_i$ \end{small}
as the consumed power of all symbols utilized to satisfy the MSE constraint $\Gamma_b \leq \gamma_b$.
 $P_a=P_t-P_c$ represents the residual power after power allocation for different subcarriers.
When the MSE constraint is met, i.e.,  $\Gamma_b \leq \gamma_b$, there exists a minimum power consumption
 $P_c$ which not only satisfying the legitimate transmission quality,
 but also has the largest residual power $P_a$ in generating AN.
 Therefore, the task of the AN-assisted scheme is to minimize the power
of all subcarriers $P_c$ under the constraint of Alice-to-Bob transmission quality.
The optimal problem can be formulated as follows:
%\begin{small}
\be
\begin{split}
min & \; \sum_{i=1}^{N_sN} p_i  \\
  s.t.& \;  \sum_{i=1}^{N_sN} \frac{\sigma^2_{z_b}}{\sigma^2_i p_i +\sigma^2_{z_b}  } \leq \gamma_b, \\
  & \; \sum_{i=1}^{N_sN}p_i \leq P_t,\\
  &\;  p_i \geq 0, i=1, \ldots, N_sN.
 \label{MINP}
 \end{split}
\ee
%\end{small}

The solution of this optimal problem can be easily obtained by CVX solvers.
After solving this restrained power allocation problem, the residual power
is  $P_a=P_t-P_c$. Alice generates AN in the time-domain with residual power $P_a$
to interfere the reception of the eavesdropper with the legitimate transmission not affected.
AN is applied to the transmission after inserting CP,
 the received symbols of Bob can be represented as
\begin{eqnarray}
\mathbf{y}_b &=&\mathbf{F}_{N_B} \mathbf{R}_{N_B}^{cp} \mathbf{H}( \mathbf{T}_{N_A}^{cp}\mathbf{F}_{N_A}^H \mathbf{W}_t \mathbf{s}+  \mathbf{z_a})+ \mathbf{z}_b \non \\
&=&  \widetilde{\mathbf{H}}  \mathbf{W}_t  \mathbf{s}+ \mathbf{F}_{N_B} \mathbf{R}_{N_B}^{cp} \mathbf{H}  \mathbf{z_a} +\mathbf{z}_b,
  \label{y_ban}
 %\mathbf{Y}_e &=&\mathbf{F}_{N_E} \mathbf{R}_{N_A}^{cp} \mathbf{G}( \mathbf{T}_{N_A}^{cp}\mathbf{F}_{N_A}^H \mathbf{W}_t \mathbf{s}+ \mathbf{Z_a}) + \mathbf{Z}_e \non \\
%&=&  \widetilde{\mathbf{G}}}\mathbf{s}+ \mathbf{F}_{N_E} \mathbf{R}_{N_A}^{cp} \mathbf{G} \mathbf{Z_a}   + \mathbf{Z}_e,
%  \label{y_ean}
\end{eqnarray}
where $\mathbf{z_a}\in \mathbb{C} ^{N_A(N+N_{cp}) \times 1} $ is the generated AN vector.
 $P_a = Tr \{\mathbb{E}\{ \mathbf{z_a} \mathbf{z_a}^H \} \} $  is the power of AN.

In this AN-assisted scheme, AN should be laid in the null space of the Alice-to-Bob channel link
so that AN will not influence Bob's reception.
Therefore, AN can be generated as
 \vspace*{-0.0 cm}
 \be \mathbf{z_a} =\mathbf{Q_a}\mathbf{d} , \label{AN}  \ee
where $\mathbf{Q_a}$ = null $( \mathbf{R}_{N_B}^{cp} \mathbf{H})$ is an $N_A (N+N_{cp})$-by-$N(N_A-N_s)+N_{cp}N_A$ matrix,
i.e.
\be \mathbf{R}_{N_B}^{cp} \mathbf{H} \mathbf{Q_a} = \mathbf{0} ,   \ee
\be \mathbf{Q_a} {\mathbf{Q_a} ^H} = \mathbf{I}_{N_A(N+N_{cp})},  \label{Q}  \ee
 $\mathbf{d}$ is $ (N(N_A-N_s)+N_{cp}N_A )\times 1 $  zero-mean complex Gaussian random vector which has the variance of ${\sigma}^2_d$.
Then, the received symbols of Bob in (\ref{y_ban}) can be rewritten as
 \vspace*{-0.0 cm}
% \begin{small}
\begin{eqnarray}
\mathbf{y}_b &=&\mathbf{F}_{N_B} \mathbf{R}_{N_B}^{cp} \mathbf{H}( \mathbf{T}_{N_A}^{cp}\mathbf{F}_{N_A}^H \mathbf{W}_t \mathbf{s}+  \mathbf{z_a})+ \mathbf{z}_b \non \\
&=&  \widetilde{\mathbf{H}}\mathbf{W}_t   \mathbf{s}+\mathbf{z}_b.
  \label{y_ban1}
\end{eqnarray}
%\end{small}

As for Eve, due to the diversity of channel characteristic, Eve cannot eliminate the
affect of AN. The received signal at Eve can be expressed as
\begin{eqnarray}
 \mathbf{y}_e &=&\mathbf{F}_{N_E} \mathbf{R}_{N_E}^{cp} \mathbf{G}( \mathbf{T}_{N_A}^{cp}\mathbf{F}_{N_A}^H \mathbf{W}_t \mathbf{s}+ \mathbf{z_a}) + \mathbf{z}_e \non \\
&=&  \widetilde{\mathbf{G}} \mathbf{W}_t  \mathbf{s}+ \mathbf{F}_{N_E} \mathbf{R}_{N_E}^{cp} \mathbf{G} \mathbf{z_a}   + \mathbf{z}_e.
  \label{y_ean}
\end{eqnarray}

Since the legitimate
channel $ \mathbf{H}$ is different from the eavesdropping channel $ \mathbf{G}$, i.e.
$ \mathbf{R}_{N_E}^{cp} \mathbf{G}\mathbf{z_a} \neq \mathbf{0} $,
 Eve cannot cancel out AN as  Bob does.
Therefore, the proposed AN-assisted scheme can further prevent Eve from eavesdropping and increase the security capacity of the  MIMO-OFDM systems.

\section{Simulation Results}
\label{sec: simulation}

%图1
  \begin{figure}[!tp]
   \begin{center}
    %\vspace{-0.3cm}
       \includegraphics[width= 3.7 in]{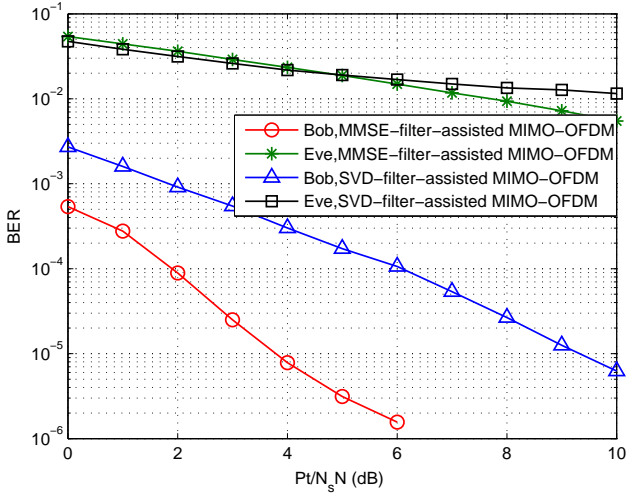}
       %\vspace{-0.3cm}
  \caption{BER versus transmit power ($N = 64$ subcarriers, length of CP $N_{cp}=16, N_A=4, N_B=N_E=N_s=2$).}
     \label{F11}
      \end{center}
   \end{figure}

\begin{center} %图2 N=128
  \begin{figure}[!tp]
  \begin{center}
    %\vspace{-0.3cm}
        \includegraphics[width= 3.7 in]{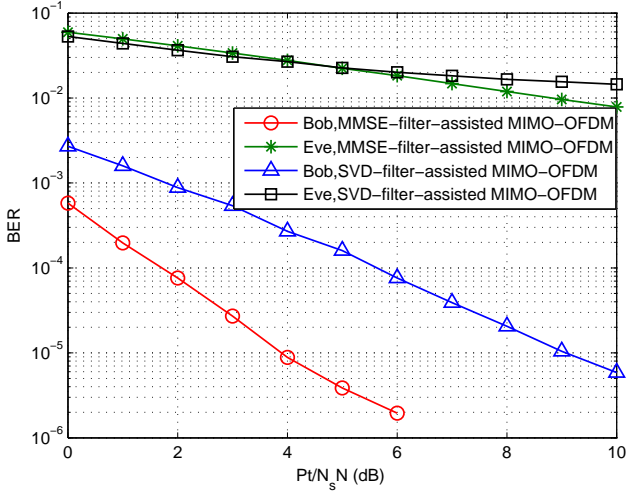}
        % \vspace{-0.3 cm}
    \caption{BER versus transmit power  ($N = 128$ subcarriers, length of CP $N_{cp}=32$, $N_A=4, N_B=N_E=N_s=2$).}
   \label{F12}
    \end{center}
   \end{figure}
\end{center}

In this section, we examine  the performance of the proposed transmit-filter-assisted and AN-assisted
secure MIMO-OFDM systems.
We evaluate the average bit-error-rate (BER)
of both Bob and Eve for various total transmit power,
the number of subcarriers, the number of antennas at Alice and MMSE constraints.
The simulation results are experienced over $10^{5}$ channel realizations.

In this simulation, Alice tries to  establish a secure transmission between herself and Bob in the
presence of Eve, using a MIMO-OFDM wireless communication system which has
$N = 64$ subcarriers and $N_{cp}=16$.
The number of antennas at Alice is $N_A=4$, and $N_B=N_E=2$, $N_s=2$.
Fig. \ref{F11} shows the BERs of Bob and Eve under different total transmit power constraints.
For comparison purpose, we compare the SVD-filter-based scheme \cite{SVD} with our proposed algorithm.
For the proposed  transmit-filter-assisted scheme,
 the BER of Bob is much lower than the
 conventional SVD-filter-based MIMO-OFDM system.
However, the BERs of Eve in both our proposed system and SVD-filter-based system
are merely the same and very high, which means that
Eve's reception has been significantly degraded.
We can also observe from Fig. \ref{F11} that the BER gap between Bob and Eve in our proposed scheme
is significantly enlarged comparing with the conventional SVD scheme, which demonstrates that our proposed
MMSE-filter-assisted MIMO-OFDM system has better security performance.

Then, the simulation results for a MIMO-OFDM system with $N = 128$ subcarriers and $N_{cp}=32$ are  shown in Fig. \ref{F12},
which has the similar performance as Fig. \ref{F11}.
The BER of Bob in MMSE-filter-assisted system
 can achieve (by all practical measures) errorless transmission at SINR 6dB, while Eve has much higher error rate.
These findings demonstrate that the proposed MMSE-filter-assisted scheme can significantly ensure the security of MIMO-OFDM systems.

 %天线数
  \begin{figure}[!tp]
  %\begin{center}
    %\vspace{-0.3cm}
        \includegraphics[width= 3.7 in]{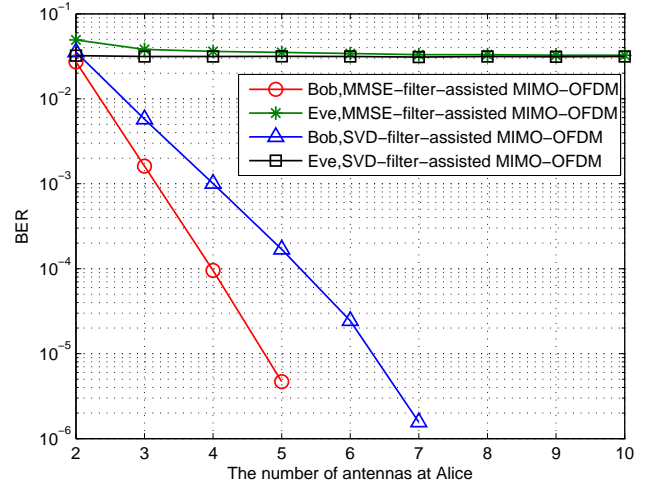}
         %\vspace{-0.3 cm}
    \caption{BER versus the numbers of antennas at Alice ($N = 64$ subcarriers, $N_{cp}=16$, $N_B=N_E=N_s=2$).}
\label{FNA}
  %  \end{center}
   \end{figure}

%  天线数
Fig. \ref{FNA} shows the BERs of Bob and Eve for various numbers of antennas at Alice.
We evaluate the number of $N_A$ from 2 to 10 and fix $N_B=N_E=N_s=2$.
The number of subcarriers $N = 64$ and the length of CP $N_{cp}=16$.
From Fig. \ref{FNA} we can notice that with the increase of the number of antennas at Alice,
the BERs of Bob will decrease to a large extent while the BERs of Eve remains unchanged.
The reason is that the larger antennas of Alice, the more degrees of freedom (DoF) to
design signals for anti-eavesdropping.
In Fig. \ref{FNA}, the BER of Bob with the proposed approach is much lower than the SVD scheme
which illustrates our proposed scheme can achieve much better performance in the MIMO-OFDM wireless communications.

\begin{figure}[!tp]
  \begin{center}
   % \vspace{-0.3cm}
        \includegraphics[width= 3.7 in]{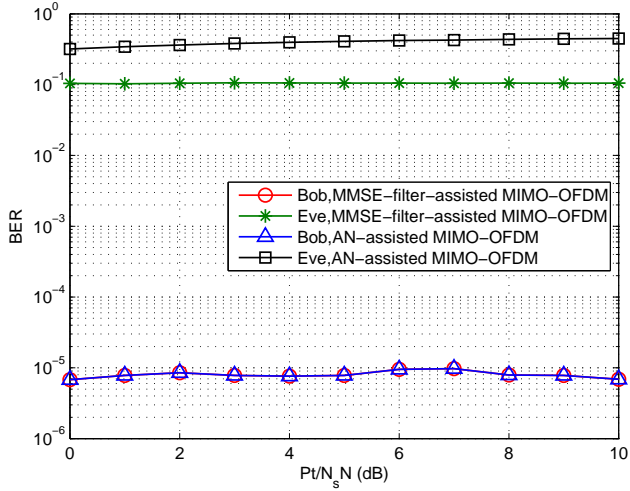}
         %\vspace{-0.3 cm}
    \caption{BER versus total transmit power
    ($N = 64$, $N_{cp}=16$, $N_A=4, N_B=N_E=N_s=2,$ MMSE constraint $\gamma_b=10$).}
\label{F21}
    \end{center}
   \end{figure}

Next, we present the performance of the AN-assisted secure MIMO-OFDM scheme in Fig. \ref{F21}. We fix the MMSE constraint as $\gamma_b =10  $ and show the BERs of Bob and Eve as a function of $P_t/N_s N$.
In Fig. \ref{F21}, the BER gap between Bob and Eve in the proposed AN-assisted secure MIMO-OFDM system is
enlarged comparing with no AN scheme.
With the MMSE constraint, the BER of Bob nearly keeps in a straight line.
Moreover, the BER of Bob remains unchanged when adding AN which illustrates that AN will be canceled out at Bob
and can further disturb the reception of Eve.
%%%%%图2 NA=10

\begin{figure}[!tp]
  %\begin{center}
    %\vspace{-0.3cm}
        \includegraphics[width= 3.7 in]{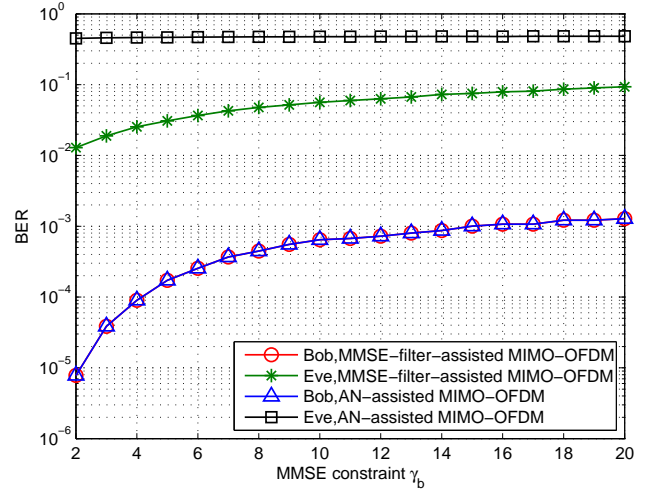}
         %\vspace{-0.3 cm}
    \caption{BER versus MMSE constraint $\gamma_b$ ($N = 64$, $N_{cp}=16$, $N_A=4, N_B=N_E=N_s=2$, transmit power $P_t=50$dB).}
\label{F3}
    %\end{center}
   \end{figure}

%图3
In Fig. \ref{F3}, the total transmit power is fixed as $P_t = 50$dB and
the horizontal axis is the MMSE constraint $\gamma_b $ of $\Gamma_b$ at Bob (i.e. $\Gamma_b \leq \gamma_b$).
Fig. \ref{F3} shows that the BER of Eve is higher when adding AN while the BER of Bob remains unchanged.
It illustrates that AN can prevent Eve from eavesdropping and maintain the security of Alice-to-Bob transmission.
Fig. \ref{F3} also shows that with the increase of $\gamma_b$, the performance of Bob will be worse
which demonstrates that our proposed MMSE-filter-assisted MIMO-OFDM scheme can influence the performance of Bob.
The numerical results show that the proposed AN-assisted secure MIMO-OFDM  scheme can achieve better security transmission than
the MIMO-OFDM system without AN.

%\vspace{-0.7 cm}
\section{Conclusion}
\label{sec: conclusion}
In this paper, we considered the problem of securing the  MIMO-OFDM wireless transmission
between the transmitter and the legitimate receiver in the presence of an eavesdropper.
We proposed two novel schemes to improve physical layer security
for MIMO-OFDM wireless communication systems.
The MMSE-filter-assisted secure MIMO-OFDM system aims to disturb the orthogonality of the received signals
at Eve to enhance the security of the Alice-to-Bob transmission.
To further improve the performance,
we proposed another AN-assisted scheme to
enhance the physical layer security in MIMO-OFDM systems.
%Due to the diversity of the legitimate channel and the wiretap channel link,
%AN can be canceled out at Bob while disturb the reception of Eve.
Simulation results shown that the proposed schemes can
significantly improve the physical layer security
for the MIMO-OFDM wireless communication systems.

%图2 NA=4

 %tu 3

%\section{Acknowledgments}
%This paper is supported by the Natural Science Foundation of China (Grant No. 61671101 and 61601080), Natural Science Foundation of Liaoning Province (Grant No. 2015020043), and the Fundamental Research Funds for the Central Universities (Grant No. DUT 15RC(3)121).
%

%\end{spacing}
\end{document}